\documentclass[10pt,conference]{IEEEtran}
\usepackage{mathrsfs}
\usepackage{amsfonts}
\usepackage{amsmath}
\usepackage{graphicx}
\usepackage{subfigure}
 
\newtheorem{theorem}{Theorem}

\newtheorem{lemma}[theorem]{Lemma}

\newcommand{\suppress}[1]{}

\newcommand{\bl}{N}
\newcommand{\spc}{{\cal S}}

\newcommand{\bX}{{\bf X}}
\newcommand{\bR}{{\bf R}}

\newcommand{\F}{\mathbb{F}_q}

\newcommand{\iden}{I}

\newcommand{\cC}{\mathcal{C}}

\newcommand{\cP}{\mathcal{P}}

\newcommand{\cN}{\mathcal{N}}

\newcommand{\negspace}{\hspace*{-2mm}}

\newcommand{\defn}{\stackrel{\triangle}{=}}

\title{Feasible alphabets for communicating the sum of sources over a network}

\author{
\authorblockN{Brijesh Kumar Rai and Bikash Kumar Dey}
\authorblockA{Department of Electrical Engineering \negspace \\
Indian Institute of Technology Bombay\negspace \\
Mumbai, India, 400 076\\
\{bkrai,bikash\}@ee.iitb.ac.in}
}

\begin{document}

\maketitle

\begin{abstract}
We consider directed acyclic {\em sum-networks} with $m$ sources
and $n$ terminals where the sources generate symbols from an arbitrary
alphabet field $F$, and the terminals need to recover the sum of
the sources over $F$. We show that
for any co-finite set of primes, there is a sum-network which is
solvable only over fields of characteristics belonging to that set.
We further construct a sum-network where a scalar solution exists over
all fields other than
the binary field $F_2$. We also show that a sum-network is
solvable over a field if and only if its reverse network is solvable
over the same field.
\end{abstract}

\section{Introduction}
\label{sec:intro}
After its introduction by the seminal work by Ahlswede et al.
\cite{AhlCLY:00}, the field of network coding has seen an explosion
of interest  and development. See, for instance, \cite{LiYC:02,koetter2,
JagSanCEEJT, medard} for some early development in the area.
The work by Dougherty et al. in \cite{dougherty1, dougherty2} are specially
relevant in the context of this paper
for the nature of results and the proof techniques.
They defined a network with specific demands of the terminals
to be scalar linear solvable (resp. vector linear solvable)
over a field $F_q$ if there exists a scalar linear network code (resp.
vector linear network code) over $F_q$ which satisfies the demands of
all the terminals.
A prime $p$ is said to
be a characteristic of a network if the network is solvable over
some finite field of characteristic $p$. They showed that for any finite
or co-finite set of primes, there exists a network where the given set
is the set of characteristics of the network.

In most of the past work, the
terminal nodes have been considered to require the recovery of all or
part of the sources' data.
A more general setup is where the terminals
require to recover some functions of the sources' data.
Recently, the problem of communicating the sum of sources to
some terminals was considered in \cite{ramamoorthy,RaiD:09}. We call
such a network as a sum-network. It was shown
in~\cite{ramamoorthy}
that if there are two sources or two terminals in the network, then the sum
of the sources can be communicated to the terminals if and only if every
source is connected to every terminal. While this condition is also
necessary for any number of sources and terminals, it may not be sufficient.
In~\cite{RaiD:09}, the authors showed that for any finite set of prime
numbers, there exists a network where the sum of the sources can be
communicated to the terminals using scalar or vector linear network coding if and only if
the characteristic of the alphabet field belongs to the given set.

It is worth mentioning that the problem of distributed function
computation in general has
been considered in different contexts in the past.
The work in \cite{tsistsiklis,gallager2,giridhar,kanoria,boyd}
is only to mention a few.

Given a multiple unicast network, its reverse network is obtained by reversing
the direction of all the links and interchanging the role of source
and destination for each source-destination pair. It is known~(\cite{riis1, dougherty3})
that a multiple unicast network is linearly solvable if and only
if its reverse network is linearly solvable. However, there are
multiple-unicast networks which are solvable by nonlinear network coding
but whose reverse networks are not solvable~(\cite{riis1, dougherty3}).

In this paper, we consider a directed acyclic network with
unit-capacity links. We prove the
following results.

\begin{itemize}
\item  For every co-finite set of prime numbers, there exists a directed
acyclic network of unit-capacity links with some sources and terminals
so that the sum of the sources can be communicated to all the terminals
using vector or scaler network coding
if and only if the characteristic of the alphabet field belongs to the
given set. This result complements the result in \cite{RaiD:09}.

\item We construct a network where the sum of the sources can be communicated
to the terminals over all fields except the binary field $F_2$.
This shows that whether the sum of the sources can be communicated
to the terminals in a network using scalar linear network coding
over a field does not depend only on the characteristic of the field.
It may also depend further on the order of the field.

\item The sum of the sources can be communicated to the terminals in a network
over some alphabet field using linear network coding if and only if the same
is true for the reverse network.
\end{itemize}

Proof techniques of this paper are similar to that in~\cite{RaiD:09}.

In Section \ref{sec:back}, we introduce the system model. The results
of this paper are presented in Section \ref{results} and Section
\ref{sec:reverse}. We conclude the
paper with a short discussion in Section \ref{disc}.

\section{System model}\label{sec:back}

A {\em sum-network} is represented by a directed acyclic graph $G= (V,E)$ where $V$ is a
finite set
denoting the vertices of the network, $E \subseteq V \times V$ is
the set of edges. Among the vertices, there are $m$ sources $s_1, s_2, \cdots, s_m \in V$,
and $n$ terminals $t_1, t_2, \cdots, t_n \in V$ in the network.
For any edge $e=(i,j)\in E$, the node $j$ will be called the head
of the edge and the node $i$ will be called the tail of the edge; and
they will be denoted as $head (e)$ and $tail (e)$ respectively.
Throughout the paper, $p$, possibly with subscripts, will denote a
positive prime integer, and $q$ will denote a power of a prime.
Let $\F$ denote the alphabet field. Each link in the network is
assumed to be capable of carrying a symbol from $\F$ in each use.
Each link is used once in every symbol interval and this time is taken
as the unit time. Each source generates one symbol from $\F$ in every
symbol interval, and each
terminal requires to recover the sum of the source symbols (over $\F$).

For any edge $e\in E$, let $Y_e \in \F$ denote the message transmitted
through $e$.
In scalar linear network coding, each node computes a linear combination
of the incoming symbols for transmission on an outgoing link. That is,
\begin{eqnarray}
Y_e & = & \sum_{e^\prime : head(e^\prime ) = tail(e)} \alpha_{e^\prime, e}
Y_{e^\prime}\label{lcode1}
\end{eqnarray}
when $tail(e)$ is not a source node.
Here $\alpha_{e^\prime, e} \in \F$ are called the local coding
coefficients. A source node computes a linear combination of some
data symbols generated at that source for transmission on an outgoing link,
that is,
\begin{eqnarray}
Y_e & = & \sum_{j : X_j \mbox{ generated at }tail(e)} \beta_{j, e}
X_{j}\label{lcode2}
\end{eqnarray}
for some $\beta_{j,e} \in \F$ if $tail(e)$ is a source node.
Since each source generates one symbol from $\F$ per unit time,
there is only one term in the summation in (\ref{lcode2}) and $\beta_{j,e}$
can be taken to be 1 without loss of generality.
The decoding operation at a terminal involves taking a linear combination
of the incoming messages to recover the required data.

In vector linear network coding, the data stream generated at each source
node is blocked in vectors of length $\bl$. The coding operations
are similar to  (\ref{lcode1}) and (\ref{lcode2}) with the difference
that, now $Y_e, Y_{e^\prime}, X_j$ are vectors from $\F^\bl$, and
$\alpha_{e^\prime, e}, \beta_{j,e}$ are matrices from $\F^{\bl \times \bl}$.
It is known that scalar linear network coding may give better throughput in some
networks than that is achievable by routing.
Vector linear network coding may give further improvement over scalar
linear network coding in some networks \cite{lehman, medard, riis}.

A sequence of nodes $(v_1, v_2, \ldots, v_l)$ is called a path, denoted
as $v_1 \rightarrow v_2 \rightarrow \cdots \rightarrow v_l$, 
if $(v_i, v_{i+1}) \in E$ for $i=1,2,\ldots, l-1$.
Given a network code on the network, $\prod_{i=1}^{l-2} \alpha_{(v_i, v_{i+1}),
(v_{i+1}, v_{i+2})}$ is called the path gain of the path
$v_1 \rightarrow v_2 \rightarrow \cdots \rightarrow v_l$.

A sum-network is said to be $N$-length vector linear solvable over $F_q$
if there is a $N$-length vector linear network code so that each
terminal recovers the sum of the $N$-length vectors over $F_q$ generated
at all the sources. Scalar linear solvability of a sum-network is defined
similarly.

\section{Results}
\label{results}

\begin{center}
\begin{figure}[h]
\centering\includegraphics[width=3.0in]{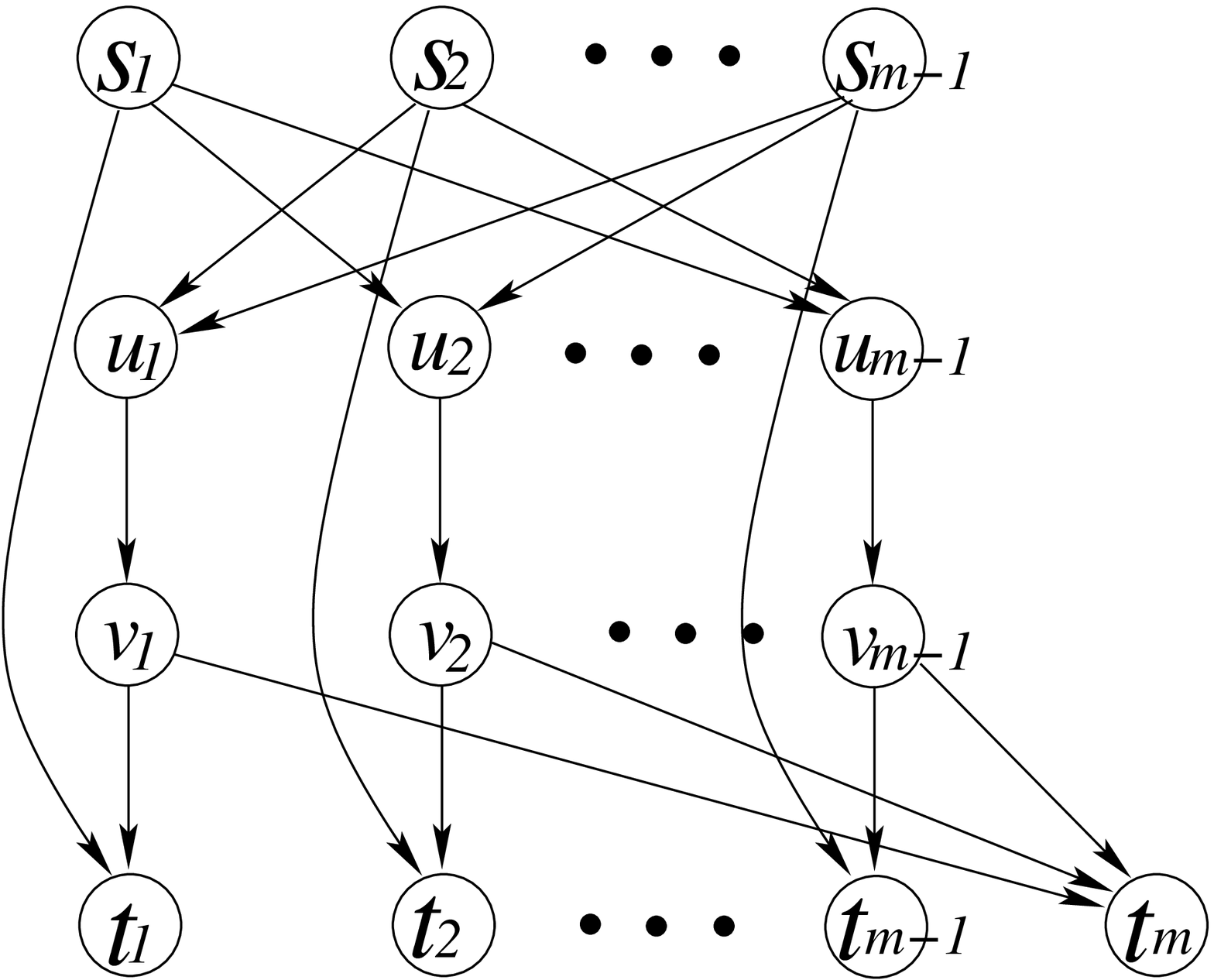} \caption{The
network $\spc_m^{*}$} \label{fig:spcm*}
\end{figure}
\end{center}

In~\cite{RaiD:09}, a special network $\spc_m$ was defined where
the sum of the sources can be communicated to the terminals using
scalar or vector linear network coding only over fields of characteristics
dividing $m-2$. For $m\geq 3$, we now define a network
$\spc_m^{*} \defn (V(\spc_m^{*}), E(\spc_m^{*}))$ which has
four layers of vertices $V(\spc_m^{*}) = S \cup U \cup V \cup T$.
The first layer of nodes are the $m-1$ source nodes $S \defn \{s_1,
s_2, \ldots, s_{m-1}\}$. The second and third layers have $m-1$
nodes each, and they are denoted as $U \defn \{u_1, u_2, \ldots,
u_{m-1}\}$ and $V \defn \{v_1, v_2, \ldots, v_{m-1}\}$ respectively.
The last layer consists of the $m$ terminal nodes $T \defn \{t_1,
t_2, \ldots, t_m\}$. For every $i=1,2,\ldots, m-1$, there is an edge
from $s_i$ to $t_i$, $u_i$ to $v_i$, $v_i$ to $t_i$, and from
$v_i$ to $t_m$. That is, $(s_i, t_i), (u_i, v_i), (v_i, t_i), (v_i, t_m)
\in E(\spc_m^{*})$ for each $i=1,2,\ldots , m-1$.
Also for every $i,j=1,2,\ldots, m-1$, $i\neq j$, there is an edge
from $s_i$ to $u_j$. So, the set of edges is given by
\begin{eqnarray}
E(\spc_m^*)  =
& \cup_{i=1}^{m-1} & \{(s_i, t_i), (u_i, v_i), (v_i, t_i), (v_i, t_m)\}\nonumber \\
& \cup & \{(s_i, u_j): i,j = 1,2,\ldots, m-1, i\neq j\} \nonumber
\end{eqnarray}
The network is shown in Fig. \ref{fig:spcm*}.

\vspace*{-4mm}

\vspace*{-7mm}
Now we present a lemma which will be used to prove one of the main
results of this paper.

\begin{lemma}
\label{lem:cofinite}
For any positive integer $N$, the network $\spc_m^{*}$ is
$N$-length vector linear solvable if and only if the characteristic
of the alphabet field does not divide $m-2$.
\end{lemma}

\begin{proof}
First we note that every source-terminal pair
in the network $\spc_m^{*}$ is connected. This is clearly a
necessary condition for being able to communicate the sum of the
source messages to each terminal node over any field.

We now prove that if it is possible to communicate the sum of the
source messages using vector linear network coding over
$\F$ to all the terminals in $\spc_m^{*}$, then the characteristic
of $\F$ must not divide $m-2$. As in (\ref{lcode1}) and 
(\ref{lcode2}), the message carried by an edge $e$ is denoted by
$Y_e$. For $i = 1,2,\ldots, m$, the message vector generated by the
source $s_i$ is denoted by $X_i \in \F^\bl$. Each terminal $t_i$
computes a linear combination $R_i$ of the received vectors.

Local coding coefficients/matrices used at different layers in the network
are denoted by different symbols for clarity. The message vectors
carried by different edges and the corresponding local coding
coefficients are as below.
\begin{subequations}\label{eq:smcode_new}
\begin{eqnarray}
Y_{(s_i, t_i)} & = & \alpha_{i,i} X_i
\mbox{ for } 1\leq i \leq m-1,\label{code1_new} \\
Y_{(s_i, u_j)} & = & \alpha_{i,j} X_i \nonumber\\
&& \hspace*{6mm} \mbox{ for } 1\leq i,j \leq m-1, i\neq j,\label{code2_new}
\\
Y_{(u_i, v_i)} & = & \mathop{\sum_{j=1}^{m-1}}_{j\neq i} \beta_{j,i} Y_{(s_j, u_i)}
\nonumber\\
&& \hspace*{15mm} \mbox{ for } 1\leq i \leq m-1,\label{code4_new}
\end{eqnarray}
\end{subequations}
\vspace*{-5mm}
\begin{subequations}\label{eq:smdecode_new}
\begin{eqnarray}
R_i & = & \gamma_{i,1} Y_{(s_i, t_i)}
+ \gamma_{i,2} Y_{(v_i, t_i)} \nonumber \\
&& \hspace*{20mm} \mbox{ for } 1\leq i \leq m-1,\label{code5_new} \\
R_m & = & \sum_{j=1}^{m-1} \gamma_{j,m} Y_{(v_j, t_m)}. \label{code6_new}
\end{eqnarray}
\end{subequations}
Here all the coding coefficients $\alpha_{i,j}, \beta_{i,j}, \gamma_{i,j}$ are $\bl \times \bl$ matrices over $\F$, and the message vectors $X_i$ and the messages carried by the links $Y_{(.,.)}$ are length-$\bl$ vectors over $\F$.

Without loss of generality (w.l.o.g.), we assume that
$ Y_{(v_i, t_i)} = Y_{(v_i, t_m)} = Y_{(u_i, v_i)}$ and 
$\alpha_{i,i} = \alpha_{i,j} = I$
for $1\leq i,j \leq m-1, i\neq j$, where $I$ denotes the $N \times N$
identity matrix.

By assumption, each terminal decodes the sum of all the source messages. That is,
\begin{eqnarray}
R_i = \sum_{j=1}^{m-1} X_j \mbox{ for } i=1,2,\ldots, m \label{eq:sum1_new}
\end{eqnarray}
for all values of $X_1, X_2, \ldots, X_{m-1} \in \F^\bl$.

From equations (\ref{eq:smcode_new}) and (\ref{eq:smdecode_new}), we have
\begin{equation}
R_i =  \mathop{\sum_{j=1}^{m-1}}_{j\neq i}\gamma_{i,2}\beta_{j,i}X_j+ \gamma_{i,1} X_i \label{eq:decode1_new}
\end{equation}
for $i=1,2,\ldots, m-1$, and
\begin{eqnarray}
R_m  &=&  \sum_{i=1}^{m-1}\gamma_{i,m}\left(\mathop{\sum_{j=1}^{m-1}}_{j\neq i}\beta_{j,i}X_j\right)\nonumber\\
&=& \sum_{j=1}^{m-1}\left(\mathop{\sum_{i=1}^{m-1}}_{i\neq j}\gamma_{i,m}\beta_{j,i}\right)X_j.\label{eq:decode2_new}
\end{eqnarray}
Since (\ref{eq:sum1_new}) is true for all values of $X_1, X_2, \ldots, X_m \in \F^\bl$,
equations (\ref{eq:decode1_new}) and (\ref{eq:decode2_new}) imply
\begin{eqnarray}
&& \gamma_{i,2}\beta_{j ,i} = \iden \mbox{ for } 1\leq i, j \leq m-1, i\neq j, \label{sol1_new} \\
&& \gamma_{i,1} = \iden \mbox{ for } 1\leq i\leq m-1, \label{sol2_new} \\
&& \mathop{\sum_{i=1}^{m-1}}_{i\neq j}\gamma_{i,m}\beta_{j,i} = \iden \mbox{ for } 1\leq j\leq m-1.  \label{sol3_new}
\end{eqnarray}
All the coding matrices in equations (\ref{sol1_new}),(\ref{sol2_new}) are invertible since the right hand side of the equations are the identity matrix. Equations (\ref{sol1_new}) imply
$\beta_{j,i} = \beta_{k,i}$ for $1\leq i, j, k\leq m-1$, $j\neq i\neq k$.
So, let us denote all the equal co-efficients $\beta_{j,i}; 1\leq j \leq m-1,
j\neq i$ by $\beta_i$. Then (\ref{sol3_new}) can be rewritten as
\begin{eqnarray}
\mathop{\sum_{i=1}^{m-1}}_{i\neq j}\gamma_{i,m}\beta_{i} = \iden \mbox{ for }
1\leq j\leq m-1.
\label{sol4_new}
\end{eqnarray}

Equation (\ref{sol4_new}) implies
\begin{eqnarray}
\gamma_{i,m}\beta_{i} = \gamma_{j,m}\beta_{j}  \mbox{ for } 1\leq i, j\leq m-1, i\neq j. \nonumber
\end{eqnarray}
Then (\ref{sol4_new}) gives
\begin{eqnarray}
(m-2)\gamma_{1,m}\beta_{1}  = \iden \nonumber \\
\Rightarrow \gamma_{1,m}\beta_{1} = (m-2)^{-1}\iden .  \label{sol6_new}
\end{eqnarray}
Equation (\ref{sol6_new}) implies that the matrix $\gamma_{1,m}\beta_{1}$
 is a diagonal matrix and all the diagonal elements are equal to $(m-2)^{-1}$.
But the inverse of $(m-2)$ exists over the alphabet field if and
only if the characteristic of the field does not divide $(m-2)$. So,
the sum of the source messages can be communicated in $\spc_m^{*}$ by $N$-length vector
linear network coding over $F_q$ only if the characteristic of $\F$ does not
divide $(m-2)$.

Now, if the characteristic of $\F$ does not divide $(m-2)$, then for any block
length $\bl$, in particular for scalar network coding for $\bl = 1$,
every coding matrix in (\ref{code1_new})-(\ref{code4_new}) can be
chosen to be the identity matrix. The terminals $t_{1},t_{2},\cdots,
t_{m-1}$ then can recover the sum of
the source messages by taking the sum of the incoming messages, i.e., by
taking $\gamma_{i,1} = \gamma_{i,2} = \iden$ for $1\leq i\leq m-1$
in (\ref{code5_new}). Terminal $t_{m}$ recovers the sum of the
source messages by taking $\gamma_{i,m}, 1\leq i\leq m-1$ in
(\ref{code6_new}) as diagonal matrices having diagonal elements as
inverse of $(m-2)$. The inverse of $(m-2)$ exists over $F_q$ because
the characteristic of $F_q$ does not divide $(m-2)$.
\end{proof}

Lemma \ref{lem:cofinite} gives the following theorem.
\begin{theorem}
\label{thm:main_cofinite}
For any finite set $\cP = \{p_1, p_2, \ldots, p_l\}$
of positive prime numbers, there exists a directed acyclic sum-network
of unit-capacity edges where for any positive integer $N$, the network
is $N$-length vector linear solvable if and
only if the characteristic of the alphabet field does not belong to $\cP$.
\end{theorem}
\begin{proof}
Consider the network $\spc_m^*$ for $m = p_1p_2\ldots p_l + 2$. This
network satisfies the condition in the theorem by Lemma \ref{lem:cofinite}.
\end{proof}

We note that the alphabet field in Theorem \ref{thm:main_cofinite} may
also be an infinite field of non-zero characteristic. In particular,
the theorem also applies to the field of rationals $F_q (X)$
over $F_q$. So, the sum-network in Theorem \ref{thm:main_cofinite}
is also solvable using linear convolutional network code over $F_q$
if and only if the characteristic of $F_q$ is not in $\cP$.

Now we define another sum-network $G_1$ with the set of vertices
$V(G_1) \defn \cup_{i=1}^3 \{s_i, u_i, v_i, t_i\}$, edges
$E(G_1) \defn \{(u_i, v_i)|i=1,2,3\}\cup \{(s_i, u_j), (v_i, t_j)|i,j=1,2,3, i\neq j\}$. The network is shown in Fig. \ref{fig:f2}. The nodes $s_1, s_2, s_3$ are
the sources and the nodes $t_1, t_2, t_3$ are the terminals in the network.
The symbols generated at the sources are denoted by $X, Z$, and $W$ respectively.
\begin{center}
\begin{figure}[h]
\centering\includegraphics[width=2.5in,height=2.5in]{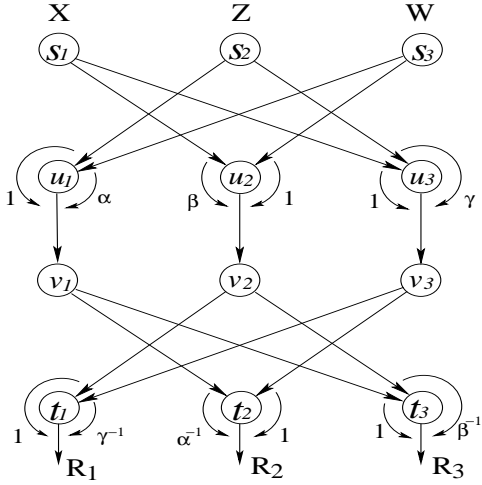}
\caption{The network $G_1$}
\label{fig:f2}
\vspace*{-4mm}
\end{figure}
\vspace*{-2mm}
The following lemma gives our second main result.
 \end{center}
\begin{lemma} The sum-network $G_1$ is scalar linear solvable over all
fields other than $F_2$.
\end{lemma}
\begin{proof}
The message vectors carried by different edges and the corresponding local coding coefficients
are as below.\\
Without loss of generality, we assume
\begin{subequations}\label{eq:bcode1_binary}
\begin{eqnarray}
Y_{(s_1, u_2)} & = & Y_{(s_1, u_3)}  =  X, \label{code1_binary}\\
Y_{(s_2, u_1)} & = & Y_{(s_2, u_3)}  =  Z ,\label{code2_binary}\\
Y_{(s_3, u_1)} & = & Y_{(s_3, u_2)}  =  W, \label{code3_binary}
\end{eqnarray}
\end{subequations}
and
\begin{subequations}\label{eq:bcode2_binary}
\begin{eqnarray}
Y_{(u_1, v_1)} & = & Y_{(s_2, u_1)}+\alpha Y_{(s_3, u_1)},\label{code4_binary}\\
Y_{(u_2, v_2)} & = & Y_{(s_3, u_2)}+\beta Y_{(s_1, u_2)}, \label{code5_binary}\\
Y_{(u_3, v_3)} & = & Y_{(s_1, u_3)}+\gamma Y_{(s_2, u_3)},
\label{code6_binary}
\end{eqnarray}
\end{subequations}
where $\alpha, \beta, \gamma \in F_q$.\\ 
Also, w.l.o.g, we assume that
\begin{subequations}\label{eq:bcode3_binary}
\begin{eqnarray}
Y_{(u_1, v_1)} & = & Y_{(v_1, t_2)} \ = \ Y_{(v_1, t_3)},\label{code7_binary}\\
Y_{(u_2, v_2)} & = & Y_{(v_2, t_1)} \ = \ Y_{(v_2, t_3)},\label{code8_binary}\\
Y_{(u_3, v_3)} & = & Y_{(v_3, t_1)} \ = \ Y_{(v_3, t_2)}.
\label{code9_binary}
\end{eqnarray}
\end{subequations}
Since there is only one path $s_2 \rightarrow u_3 \rightarrow v_3 \rightarrow
t_1$ from source $s_2$ to terminal $t_1$
and also one path $s_3 \rightarrow u_2 \rightarrow v_2 \rightarrow t_1$
from source $s_3$ to $t_1$ with path gains $\gamma$ and
$1$ respectively, the recovered symbol $R_1$ at $t_1$ must be
\begin{subequations}\label{eq:bdecode1_binary}
\begin{eqnarray}
R_1  &  = &  Y_{(v_2, t_1)} + \gamma^{-1}Y_{(v_3,
t_1)} \label{decode1_binary}.
\end{eqnarray}
\end{subequations}
\addtocounter{equation}{-1}
Similarly, the recovered symbols $R_1$ and $R_2$ should be
\begin{subequations}\label{eq:bdecode1_binary}
\addtocounter{equation}{+1}
\begin{eqnarray}
R_2  &  = & Y_{(v_3, t_2)} + \alpha^{-1}Y_{(v_1,
t_2)},\label{decode2_binary}\\
 R_3 & = & Y_{(v_1, t_3)} +
\beta^{-1}Y_{(v_2, t_3)}\label{decode3_binary}.
\end{eqnarray}
\end{subequations}
The coding coefficients are depicted in Fig. \ref{fig:f2} for clarity.

From equations (\ref{eq:bcode1_binary}), (\ref{eq:bcode2_binary}), (\ref{eq:bcode3_binary}) and (\ref{eq:bdecode1_binary}) it follows that
\begin{subequations}\label{eq:bcondition1_binary} 
\begin{eqnarray}
R_1 & = & (\beta + \gamma^{-1})X+Z+W,  \label{condition1_binary}\\
R_2 & = & X+(\gamma + \alpha^{-1})Z+W, \label{condition2_binary}\\
R_3 & = & X+Z+(\alpha + \beta^{-1})W. \label{condition3_binary}
\end{eqnarray}
\end{subequations}
Note that equation (\ref{eq:bcondition1_binary}) requires that the coding coefficients $\alpha$, $\beta$ and $\gamma$ be 
non-zero. This requirement can also be seen as natural since if any of these coefficients is zero, then a particular source-terminal pair will be disconnected.\\
Since all the terminals must recover the sum of the source messages, i.e., $R_1=R_2=R_3=X+Z+W$, we have
\begin{subequations}\label{eq:bcondition2_binary} 
\begin{eqnarray}
\beta + \gamma^{-1} & = & 1, \label{condition4_binary}\\
\gamma + \alpha^{-1} & = & 1,\label{condition5_binary}\\
\alpha + \beta^{-1} & = & 1. \label{condition6_binary}
\end{eqnarray}
\end{subequations}
Now, over the binary field the values of $\alpha$, $\beta$ and
$\gamma$ must all be $1$. Putting $\alpha=\beta=\gamma=1$ in equation
(\ref{eq:bcondition2_binary}), we have
$1  =  0$.
This gives a contradiction. So, it is not possible to communicate the sum
of the sources to the terminals in this network over the binary field $F_2$ using scalar linear network
coding.

Now, we consider any other finite field $F_q$ ($q \neq 2$). We show
that over $F_q$, the conditions in equation (\ref{eq:bcondition2_binary})
are satisfied for some choice of $\alpha, \beta$, and $\gamma$.

Since $q > 2$, let $\alpha \in F_q$ be any element other
than $0$ and $1$. Also, take $\gamma = 1 - \alpha^{-1}$
and $\beta = (1-\alpha )^{-1}$. Clearly, they satisfy
(\ref{condition4_binary})-(\ref{condition6_binary}). Hence, it is possible to communicate
the sum of the source messages to the terminals over $F_q$. 
\end{proof}

It is worth noting that
though the sum can not be communicated in this network by scalar network coding
over $F_2$, it is possible to do so by vector network coding
over $F_2$ using any block length $N > 1$. This follows because
it is possible to communicate the sum over the extension field $F_{2^N}$
using scalar network coding.

\section{Reversibility of networks}
\label{sec:reverse}
Given a sum-network (recall the definition from Sec. \ref{sec:intro})
$\cN$, its reverse network $\cN'$ is defined to be the network
with the same set of vertices, the edges reversed, and the role
of sources and terminals interchanged. It should be noted that since
$\cN$ may have unequal number of sources and terminals, the number
of sources (resp. terminals) in $\cN$ and that in $\cN'$ may be
different. For example, the reverse network $\spc_m^{*'}$
of $\spc_m^*$ has $m$ sources and $m-1$ terminals and so the problem
in $\spc_m^{*'}$ is to communicate the sum of the source messages
(say, $Y_1, \ldots, Y_m$) to the $m-1$ terminals.
In this section, we show that for any sum-network $\cN$ and any alphabet
field $F_q$, the sum-network $\cN$ is $N$-length vector linear solvable over $F_q$ if and only
if its reverse network $\cN'$ is $N$-length vector linear solvable over $F_q$.

\begin{center}
\begin{figure}[h]
\centering\includegraphics[width=3.0in]{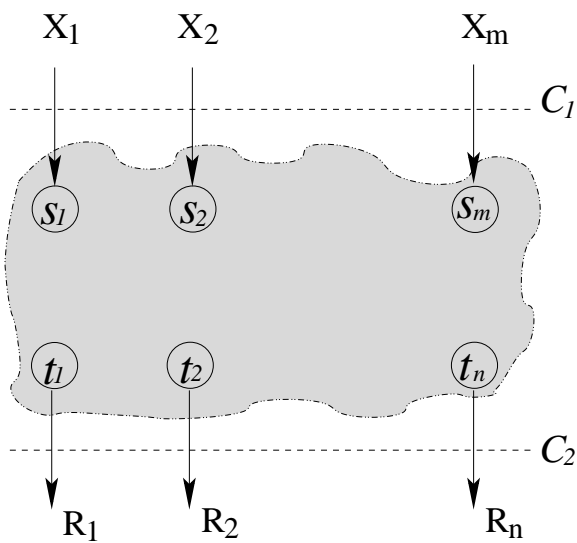} \caption{A
generic sum-network } \label{fig:sumnet}
\end{figure}
\end{center}

Consider a generic sum-network $\cN$ depicted in Fig. \ref{fig:sumnet}.
Consider the cuts $\cC_1$ and $\cC_2$ shown in the figure. We call these
cuts, the {\em source-cut} and the {\em terminal-cut} of the sum-network respectively.
The transfer function from $\cC_1$ to $\cC_2$ is defined to be
the $m\times n$ matrix $T$ over $F_q$ which relates the vectors $\bX = (X_1, X_2, \ldots,
X_m)$ and $\bR = (R_1, R_2, \ldots, R_n)$ as
\begin{eqnarray}
\bR = \bX T. \nonumber
\end{eqnarray}
In case of $N$-length vector linear coding, $X_1, X_2, \ldots, X_m, R_1, R_2,
\ldots, R_n \in F_q^N$, $\bX \in F_q^{mN}$, and $\bR \in F_q^{nN}$.
The transfer matrix is a $mN \times nN$ matrix which is easier viewed
as an $m\times n$ matrix of $N\times N$ blocks.
The $(i,j)$-th element (`block' for vector linear coding) of the
transfer matrix is the sum of the path gains of all paths from
$X_i$ to $R_j$. The following lemma follows directly.

\begin{lemma}
A sum-network $\cN$ is $N$-length vector linear solvable
if and only if there is an $N$-length vector linear network code so that
each element/block of the transfer matrix from the source-cut
to the terminal-cut is the $N\times N$
identity matrix.
\end{lemma}

Now consider the reverse network $\cN'$ of $\cN$. Let us denote the
source symbols in $\cN'$ as $Y_1, Y_2, \ldots, Y_n$ and the recovered
symbols at the terminals as $R_1^\prime, R_2^\prime, \ldots, R_m^\prime$.
Let us denote the network coding
coefficients of $\cN'$ by $\beta_{e, e'}$ for any two edges
$e, e' \in E(\cN')$ so that $head (e) = tail (e')$.
Let us denote the edge in $\cN'$ obtained
by reversing the edge $e \in E(\cN)$ by $\tilde{e}$. 
Clearly, there is a $1-1$ correspondence between the paths in $\cN$ and
the paths in $\cN'$. So, if there is a $N$-length vector linear network
code over $F_q$ which solves the sum-network $\cN$,
then $\cN'$ will also be $N$-length vector linear solvable over $F_q$
if there is an $N$-length vector linear network code for $\cN'$
which results in the same path gain for each path  in $\cN'$ as that
for the corresponding path in $\cN$. In that case, the transfer matrix from the
source-cut to the terminal-cut in $\cN'$ for that network code will
be the transpose of the transfer matrix for the network code for $\cN$.
Now, suppose $\{\alpha_{e,e'} \mid e,e' \in E(\cN), head (e) = tail (e')\}$
is the network code which solves the sum-network $\cN$. Then
clearly the network code $\{\beta_{\tilde{e'}, \tilde{e}}
\defn \alpha_{e,e'} \mid e,e' \in E(\cN),  head (e) = tail (e')\}$
results in a transfer matrix with all blocks as $I$ for $\cN'$, and thus solves
the sum-network $\cN'$. So, we have our final result:

\begin{theorem}
A sum-network $\cN$ is $N$-length vector linear solvable over $F_q$
if and only if its reverse network $\cN'$ is also $N$-length vector
linear solvable over $F_q$.
\end{theorem}

\section{Discussion}
\label{disc}
We have presented some results on communicating the sum of source messages
to a set of terminals. It was shown in \cite{dougherty2} that
there is a $1-1$ correspondence between systems of polynomial equations
and networks. This is a key result which implies existence of networks
with arbitrary finite or co-finite characteristic set.
Though sum-networks have very specific demands by
the terminal nodes and thus are more restricted as a class, a
complete characterization of the systems of polynomial equations
which have equivalent sum-networks is not yet known.
Investigation in this direction is in progress.

\section{Acknowledgments}
This work was supported in part by Tata Tele-services IIT Bombay Center of Excellence in Telecomm (TICET) and Bharti Centre for Communication.

\bibliographystyle{unsrt}
\bibliography{ref,sid,paper}

\end{document}